\documentclass{article} %
\usepackage[preprint]{colm2026_conference}

\usepackage{microtype}
\usepackage{hyperref}
\usepackage{url}
\usepackage{booktabs}
\usepackage{caption}
\usepackage{xcolor}
\usepackage{mdframed}
\usepackage[pass]{geometry}
\usepackage{subcaption}
\usepackage{graphicx}
\usepackage{amsmath}
\usepackage{amssymb}
\usepackage{enumitem}
\usepackage{makecell}
\usepackage{multirow}

\newif\iffiguresinmain
\figuresinmaintrue

\definecolor{darkblue}{rgb}{0, 0, 0.5}
\definecolor{lightgray}{RGB}{240,240,240}
\newmdenv[
  backgroundcolor=lightgray,
  linecolor=lightgray,
  innerleftmargin=12pt,
  innerrightmargin=12pt,
  innertopmargin=10pt,
  innerbottommargin=10pt,
  skipabove=6pt,
  skipbelow=6pt
]{graybox}

\hypersetup{colorlinks=true, citecolor=darkblue, linkcolor=darkblue, urlcolor=darkblue}

\title{Faster Completion, Less Learning: Generative AI Reduced Study Time on Math Problems and the Knowledge They Build}

\author{Sina Rismanchian\thanks{Equal contribution; co-first authors. Corresponding author: \texttt{srismanc@uci.edu}.} \\
University of California, Irvine \\
{\small \texttt{srismanc@uci.edu}} \\
\And
Hasan Uzun\footnotemark[1] \\
McGraw Hill \\
{\small \texttt{hasan.uzun@mheducation.com}} \\
\And
Jeffrey Matayoshi \\
McGraw Hill \\
{\small \texttt{jeffrey.matayoshi@mheducation.com}} \\
\And
Eric Cosyn \\
McGraw Hill \\
{\small \texttt{eric.cosyn@mheducation.com}} \\
\And
Eyad Kurd-Misto \\
McGraw Hill \\
{\small \texttt{eyad.kurd-misto@mheducation.com}} \\
}

\renewcommand{\normalsize}{\fontsize{10pt}{13pt}\selectfont}
\normalsize
\begin{document}

\ifcolmfinal
\fi

\maketitle

\begin{abstract}
How much have students' ordinary learning processes shifted in response to generative AI, and how does that affect their durable learning outcomes? Self-report surveys show little change, while small-scale behavioral studies report widespread AI use without the scale or duration to measure learning consequences. We address both questions using a ten-year panel of $3.2$ million ALEKS learning interactions for investigating time-on-task, complemented by $12.2$ million ALEKS PPL placement-assessment response times for examining proctoring and learning outcomes, with a quasi-experimental design exploiting variation in tasks that are more susceptible to AI (text-based word problems) and less susceptible to AI (interactive graph-based problems). Learning time on AI-susceptible problems declines $2.8\%$ per quarter among college students after ChatGPT's release, cumulating to $26.9\%$ over eleven quarters; high-schoolers show $31.3\%$, middle-schoolers $9.0\%$, and Grade 5 students no detectable change. Among college students, the post-ChatGPT divergence vanishes entirely under proctoring, ruling out broad efficiency gains as the likely explanation. Logistic fixed-effects models on randomly assigned proctored retention items yield a $25\%$ cumulative decline in odds of correct response; the same estimator on non-proctored assessment produces a large opposite-signed increase --- inconsistent with any platform, cohort, or curriculum explanation. These results are among the first large-scale behavioral and outcome evidence that generative AI has altered how students study and the knowledge they build --- the population-level indicator of \emph{cognitive surrender}, with direct implications for educational research, assessment governance, and AI policy.
\end{abstract}

\newmdenv[
  topline=true,
  bottomline=true,
  rightline=true,
  leftline=true,
  linewidth=1pt,
  linecolor=darkblue,
  backgroundcolor=gray!4,
  innerleftmargin=10pt,
  innerrightmargin=10pt,
  innertopmargin=8pt,
  innerbottommargin=8pt,
  skipabove=0pt,
  skipbelow=0pt
]{sigbox}

\begin{sigbox}
{\bfseries\color{darkblue} Significance Statement}\\[2pt]
Whether generative AI has changed how students study and what they learn has been hard to answer: surveys show little change, and small experiments cannot detect longitudinal shifts. Using $3.2$ million mathematics learning interactions over a decade, we document a large post-ChatGPT decline in time spent on AI-susceptible problems which is absent under proctoring. The shift carries a durable learning cost: on randomly assigned proctored retention items, odds of correct response fall by $25\%$, while the same estimator on non-proctored assessment yields a large opposite-signed increase --- a reversal that is unlikely for non-AI mechanism to produce.
\end{sigbox}
\clearpage

\section{Introduction}

The arrival of ChatGPT and generative AI chatbots \citep{reich2025future} placed a capable school-level mathematics problem-solver \citep{bubeck2023sparks} in the hands of millions of students with essentially zero marginal cost. It is widely assumed that this shift has changed how students study, nonetheless, reliable evidence on how much student behavior has actually moved is scarce. Large-scale self-report surveys of secondary and postsecondary students in the years following ChatGPT's release find little to no change in the rate at which high school students report using AI for academically dishonest behaviors relative to pre-ChatGPT baselines \citep{lee2024cheating,chen2026cheating}. This stability is at odds with widespread intuitions about AI adoption, and perhaps reflects limitations of self-report: social-desirability bias, normative ambiguity about what counts as AI misuse, and a lack of shared vocabulary for describing the increasingly fine-grained ways students interact with AI tools \citep{mccabe1993academic}. Other survey-based measurements that use sophisticated methods to mitigate social desirability bias find that 20\% to 34\% of college-level students report using AI to cheat \citep{nguyen2024unmasking, reiter2025student}; however, it remains unclear whether these respondents engaged in such behavior only once or did so repeatedly. 
On the other hand, small-scale behavioral studies of AI use in instructional contexts have shown widespread \textit{concerning AI usage} among undergradute students but are limited by sample sizes and short timeframes \citep{rismanchianartificial}. The net result is that educators, administrators, and researchers have been formulating AI policy largely in the absence of quantitative evidence about the magnitude of behavioral change that has actually occurred.


We address this gap using large-scale behavioral trace data from ALEKS, a widely deployed adaptive mathematics learning and assessment platform serving more than four million students annually. For the behavioral analyses, we draw on two time-based data sources observed over a ten-year window spanning the pre- and post-ChatGPT periods: $3.2$ million ALEKS learning interactions for the time-on-task analysis and $12.2$ million ALEKS PPL placement-assessment response-time observations for the proctoring analysis. Unlike self-reports, interaction logs capture students' actual behavior: how long they spend on each problem, which problem formats they accelerate on, and whether those patterns persist under supervision. Our primary behavioral outcome is time-on-task, a basic signal of engagement with the material \citep{chickering1987seven}. The central analytical lever is within-curriculum variation in AI susceptibility. \emph{AI-susceptible} problems are text-based word problems (e.g., proportional reasoning, rate and mixture, algebraic story problems) whose entire informational content fits in a text prompt. A student can copy the problem verbatim into an AI chatbot and then copy the final answer into the answer field within seconds. 
\emph{AI-resistant} problems are graph-based problems that require visually interpreting a plot and interactively manipulating a graphical widget inside ALEKS. There is no text to copy; the student must engage with the interface.
This is the difference that drives our identification strategy. 

The intuition is that if student are not using AI, then time-on-task trends should be similar across AI-susceptible and AI-resistant problems. If students are using AI to bypass cognitive engagement on AI-susceptible problems, then we should see a divergence in time-on-task trends between the two problem types following ChatGPT's release where time-on-task on AI-susceptible problems falls relative to AI-resistant problems. Similarly, if the divergence is driven by AI use, then it should disappear in proctored settings where AI access is restricted and the risk of detection is high.
As such, this differential susceptibility creates a within-student, within-quarter comparison between AI-susceptible and AI-resistant problems, helping absorb platform-wide trends, cohort differences, and calendar-time shocks.

Having established whether student behavior has shifted, we then ask whether that shift matters for learning outcomes. To do so, we complement the behavioral analyses with $6.7$ thousand ALEKS PPL retention observations that measure whether students retain the relevant knowledge under testing conditions. This retention analysis allows us to test whether reduced engagement with AI-susceptible problems is associated with weaker durable knowledge.
The cognitive science of active learning \citep{chi2014icap,koedinger2012knowledge} posits that active cognitive engagement at study time produces deeper understanding, higher mastery, and more durable retention than passive processing --- and recent randomized experimental evidence supports this prediction in the context of generative AI. \citet{bastani2025generative} showed that unrestricted GPT-4 access during mathematics practice improved assisted performance by $48\%$ but reduced unassisted exam scores by $17\%$ in a single-semester experiment with approximately $1{,}000$ students. Whether this performance reversal scales to millions of students over multi-year timeframes in authentic learning environments --- without any experimental manipulation --- is an open question that only large-scale observational data can answer.

We therefore investigate four research questions (RQ) in sequence. \emph{RQ1}, is there evidence in large-scale behavioral trace data that student study time has shifted on AI-susceptible problems following generative AI chatbots' proliferation, and if so, how large is the shift and how does it vary by student age? \emph{RQ2}, does the behavioral shift disappear in proctored assessment settings where AI access is restricted --- a falsification test that distinguishes AI-use-driven from general-efficiency-driven explanations? \emph{RQ3}, does the behavioral shift have measurable consequences for durable knowledge, as indexed by retention performance on the same problem types under conditions that restrict AI at the point of testing? \emph{RQ4}, does a second falsification by applying the same retention estimator to non-proctored assessments produce the opposite-signed result predicted by an AI-use mechanism but by no other candidate explanation?

Our answers, in order, are: (1) yes and the shift is substantial: cumulative learning-time reductions of $26.9\%$ in college and $31.3\%$ in high school over eleven post-ChatGPT quarters, concentrated in AI-susceptible formats, with only a small effect in middle school ($-9.0\%$) and a null effect in Grade 5, an age gradient consistent with autonomous AI access; (2) yes, the divergence vanishes entirely in response times under proctored assessment; (3) yes, retention declines by approximately $25\%$ cumulatively on randomly assigned AI-susceptible items under proctored conditions; and (4) yes, the same logistic estimator yields an $85\%$ opposite-signed increase in non-proctored conditions, a behavioral reversal inconsistent with any alternative explanation based on curriculum change, platform evolution, or cohort composition.

\section{Results}

\subsection{Behavioral Identification Strategy}

We exploit within-curriculum variation in susceptibility to AI assistance as the source of quasi-experimental identification. AI-susceptible topics consist of text-based word problems such as algebraic word problems whose full informational content can be transcribed into a large language model prompt in seconds. AI-resistant topics consist of graph-based and plot-based problems requiring visual interpretation of graphical displays and interactive manipulation of plots within the ALEKS platform interface. This visual-interactive structure cannot be replicated in a text prompt, making these problems substantially more resistant to external AI assistance. Representative examples from each class are shown in \iffiguresinmain Figure~\ref{fig:example_problems}\fi.

\iffiguresinmain
\begin{figure}[ht]
  \centering
  \begin{subfigure}{0.48\textwidth}
    \includegraphics[width=0.95\textwidth, trim=3cm 0cm 0 0]{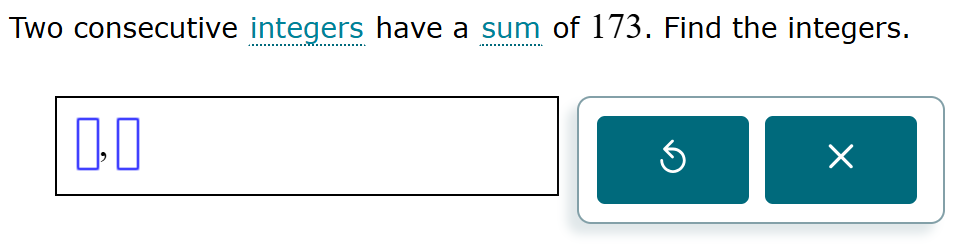}
    \caption{}
  \end{subfigure}
  \begin{subfigure}{0.48\textwidth}
    \includegraphics[width=0.9\textwidth]{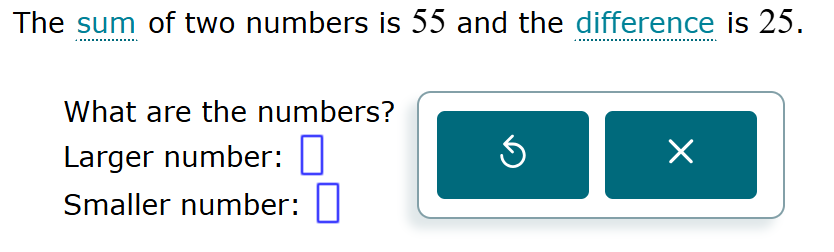}
    \caption{}
  \end{subfigure}
  \begin{subfigure}{0.48\textwidth}
    \includegraphics[width=\textwidth]{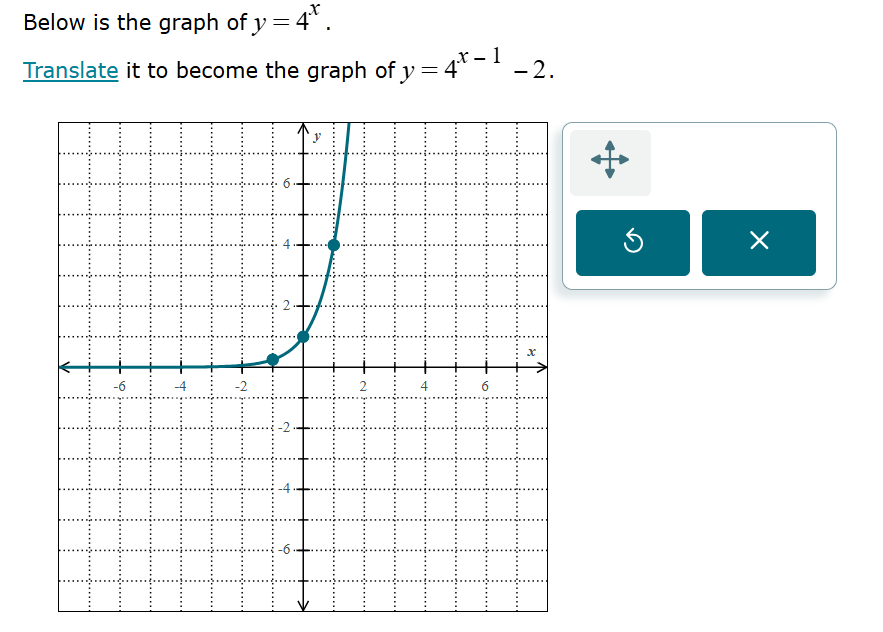}
    \caption{}
  \end{subfigure}
  \begin{subfigure}{0.48\textwidth}
    \includegraphics[width=\textwidth]{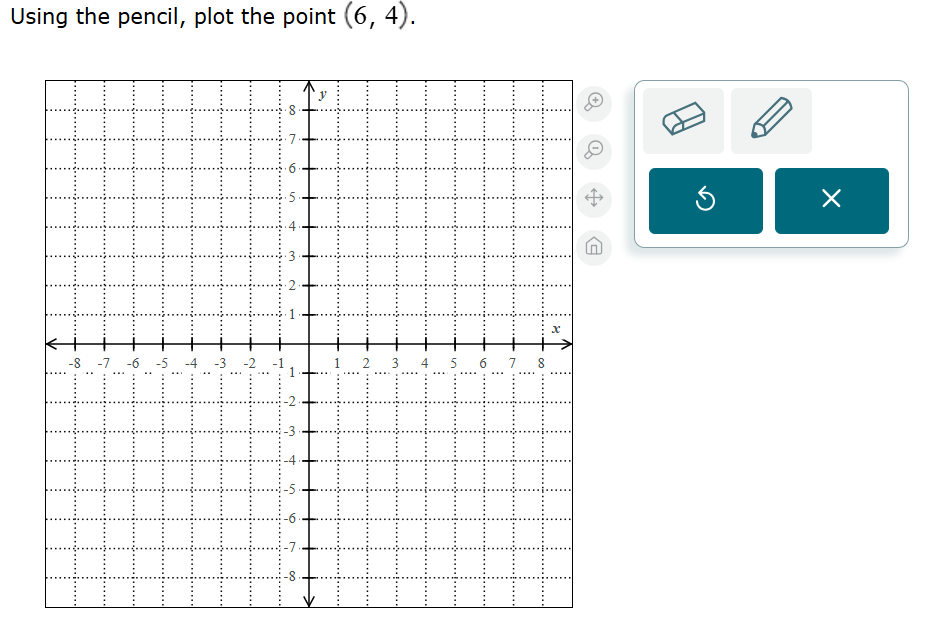}
    \caption{}
  \end{subfigure}
  \caption{\textbf{Examples of AI-susceptible and AI-resistant ALEKS items.} AI-susceptible items (a \& b) are text-based word problems whose full informational content can be transcribed into a natural-language prompt and solved by a large language model in seconds. AI-resistant items (c \& d) require visual interpretation of graphical displays and interactive manipulation of plot widgets embedded in the ALEKS interface, making them less amenable to copy-paste transfer between ALEKS and an AI chatbot.}
  \label{fig:example_problems}
\end{figure}
\fi

The learning-time analysis (RQ1) draws on $3{,}197{,}803$ ALEKS learning interactions across mathematics courses from Grade 5 through College Algebra, covering the period 2015 summer quarter (Q3) through 2025 Q3. Descriptive statistics by age group are provided in Table~\ref{tab:descriptives}. The proctoring contrast (RQ2) uses time-on-task data from ALEKS PPL, and the retention analyses (RQ3, RQ4) use learning outcomes from a separate ALEKS PPL placement-assessment panel described in the Section~\ref{ppl_data}.

\begin{table}[ht]
\centering
\small
\caption{\textbf{Descriptive statistics of the number of topics, number of attempts, average of log(time), SD(log(time)), earliest, and latest datapoints by age group.} Topic counts here include all topics in the raw extract; the SI Appendix robustness analyses (R5--R8) restrict to topics with at least $20$ recorded attempts per item--quarter cell to ensure stable cell-level means, which reduces the topic count to $281$ for College and $381$ for High School. The AI-susceptible vs.\ AI-resistant breakdown of these topic counts is reported in SI Appendix, Table~\ref{tab:topic_balance}.}
\label{tab:descriptives}
\begin{tabular}{lrrrrcc}
\toprule
Age Group & \makecell[r]{Number\\of topics} & \makecell[r]{Number of\\attempts} & \makecell[r]{Average of\\Log(Time)} & \makecell[c]{SD(Log(Time))} & \makecell[c]{Earliest\\Attempt Date} & \makecell[c]{Latest\\Attempt Date} \\
\midrule
Grade 5 & $153$ & $277{,}900$ & $5.48$ & $0.877$ & 2015 Q3 & 2025 Q3 \\
Middle School & $517$ & $761{,}190$ & $5.45$ & $0.842$ & 2015 Q3 & 2025 Q3 \\
High School & $499$ & $923{,}604$ & $5.63$ & $0.908$ & 2015 Q3 & 2025 Q3 \\
College & $334$ & $1{,}235{,}109$ & $6.12$ & $0.927$ & 2015 Q3 & 2025 Q3 \\
\bottomrule
\end{tabular}
\end{table}

The unit of analysis is the topic $\times$ calendar quarter. The dependent variable is the average log learning time for each topic in each quarter. We estimate
\begin{equation}
\bar{Y}_{j,q} \;=\; \delta_j + \tau_q + \beta\,(AISusceptible_j \times \text{TimeAfter}_q) + u_{j,q},
\label{eq:ramp}
\end{equation}
where $\bar{Y}_{j,q}$ is average log learning time for topic $j$ in quarter $q$, $\delta_j$ denotes topic fixed effects controlling for time-invariant differences in difficulty and format, $\tau_q$ denotes quarter fixed effects absorbing seasonality and platform-wide trends, and $\text{TimeAfter}_q$ equals zero before ChatGPT's release (Fall 2022) and increases linearly thereafter. This linear ramp specification captures gradual adoption dynamics consistent with technology diffusion theory \citep{rogers2003diffusion}, rather than an immediate behavioral discontinuity. Observations are weighted by the number of attempts per topic--quarter cell; standard errors are clustered at the topic level. We separately estimate event-study models with quarter-by-quarter interactions to visualize dynamic effects and assess parallel pre-trends.

For the proctoring analysis, we use a separate dataset from ALEKS PPL (Placement, Preparation, and Learning), a postsecondary mathematics placement assessment deployed at more than $400$ United States colleges and universities as the high-stakes gatekeeper into credit-bearing College Algebra and other first college mathematics courses (see Methods, \S\emph{Data}, for scale, stakes, and time span). ALEKS PPL is administered under varying supervision conditions, and we estimate separate ramp models by condition --- in-person proctored versus fully unproctored --- using an identical topic-based quasi-experimental design.

For the retention analysis, we apply logistic item and quarter fixed-effects models to binary correct/incorrect responses on ALEKS PPL retention questions administered in proctored settings:
\begin{equation}
\log\!\left(\frac{P(\text{correct}_{i,j,q})}{1 - P(\text{correct}_{i,j,q})}\right) = \delta_j + \tau_q + \beta\,(AISusceptible_j \times \text{TimeAfter}_q) + \epsilon_{i,j,q}.
\label{eq:retention}
\end{equation}
As the majority of assessment items are selected using the adaptive knowledge assessment algorithm, to remove the confounding effect of the adaptive selection of items, we restrict our analysis to ``extra'' retention questions (i.e., items assigned independently of the platform's adaptive algorithm).

\subsection{RQ1: Learning Time Declines After ChatGPT, Concentrated in AI-Susceptible Problems}

\begin{figure}[ht]
\centering
\includegraphics[
    width=0.99\linewidth,
    trim=0 2cm 0 0cm,  
    clip
]{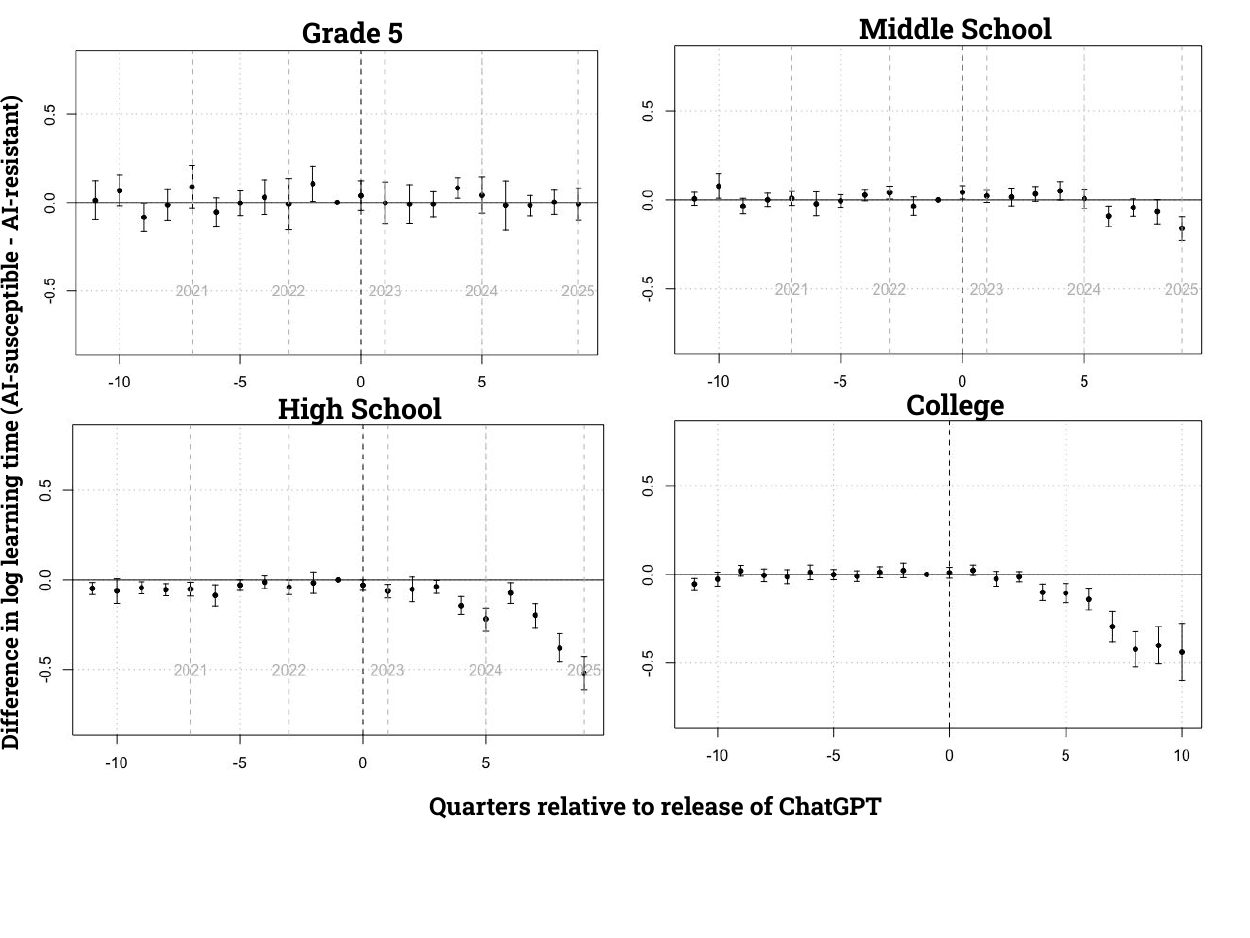}
\caption{\textbf{After ChatGPT's release, older students spent sharply less time on word problems (AI-susceptible) relative to graph problems (AI-resistant).} Each point is the estimated gap in average log learning time between AI-susceptible word problems and AI-resistant graph problems, relative to their pre-ChatGPT difference. Pre-ChatGPT coefficients are small in magnitude across all age groups; After ChatGPT's release, a growing divergence emerges for high school and college students. Error bars are $95\%$ cluster-robust confidence intervals.}
\label{fig:eventstudy_age}
\end{figure}

Figure~\ref{fig:eventstudy_age} shows event-study estimates of the differential trend in learning time between AI-susceptible and AI-resistant topics by age group.

Prior to ChatGPT's release, pre-period event-study coefficients are small in magnitude in every age group. In the Grade 5 and Middle School subsets they are statistically indistinguishable from zero. In the College and High School learning-time subsets, the pre-trend is a small \emph{positive} drift (College: $+0.0071$ log-time per quarter, $p<0.001$; HS: $+0.0046$, $p=0.003$; SI Appendix, Tables~\ref{tab:R3_placebo},~\ref{tab:R4_trends}) as AI-susceptible word problems were already becoming relatively \emph{slower} than AI-resistant graph problems before ChatGPT. Because this drift is opposite in sign to the post-ChatGPT effect, it makes the unadjusted ramp estimates we report below conservative; the trend-adjusted specification reported in SI Appendix Section~\ref{app:R9} yields a larger (more negative) post-ChatGPT estimate than the unadjusted specification. The proctored-PPL and retention specifications, in contrast, satisfy the parallel-trends assumption directly (joint Wald $p > 0.3$ in each; SI Appendix, Table~\ref{tab:R4_trends}). Following ChatGPT's release, a clear and growing divergence emerges for high school and college students.

Ramp estimates are reported in Table~\ref{tab:ramp_age}. For college-level topics, learning time for word problems declines by $2.80\%$ per quarter relative to graph problems ($\beta = -0.0284$,$p < 0.001$; per-quarter percent computed as $(\exp(\beta)-1)\times 100$), accumulating to a $26.9\%$ cumulative reduction over eleven post-ChatGPT quarters (CI: $-34.0\%$, $-19.0\%$; SI Appendix, Table~\ref{tab:R6_boot}). High school topics show a larger effect of $3.35\%$ per quarter ($\beta = -0.0341$, $p < 0.001$), corresponding to a $31.3\%$ cumulative decline (CI: $-37.3\%$, $-25.4\%$). Middle school courses show a smaller but statistically significant effect of $0.86\%$ per quarter ($\beta = -0.0086$, $p < 0.01$), accumulating to $9.0\%$. Grade 5 courses show no detectable change ($\beta = -0.0012$, $p = 0.79$). Randomization inference based on $1{,}000$ placebo permutations of the AI-susceptibility label yields $p < 0.001$ for every non-null subset, confirming that cluster-robust asymptotics are not inflating significance (SI Appendix, Table~\ref{tab:R5_perm}). Estimates are stable in sign and within roughly $30\%$ of baseline magnitude when the treatment break is moved across the four quarters spanning 2022-Q4 through 2023-Q3 (SI Appendix, Table~\ref{tab:R2_break}).

\begin{table}[ht]
\centering
\small
\caption{\textbf{Ramp estimates by age group.} Per-quarter slope coefficients $\beta$ from Equation~\ref{eq:ramp} for the differential ramp (AI-susceptible $\times$ TimeAfter) on log learning time, by age group. Standard errors clustered at the topic level. Cumulative effect over $K=11$ post-ChatGPT quarters is reported as $(\exp(K\beta)-1)\times 100\%$. Bootstrap $95\%$ CIs are from $500$ cluster bootstrap replicates.}
\label{tab:ramp_age}
\begin{tabular}{lrrrrr}
\toprule
Age group & $\beta$ & SE & $p$ & Cumulative $\%$ & Bootstrap $95\%$ CI \\
\midrule
Grade 5 & $-0.0012$ & $0.0046$ & $0.79$ & $-1.3$ & $(-8.8, +6.7)$ \\
Middle school & $-0.0086$ & $0.0033$ & $0.009$ & $-9.0$ & $(-14.9, -2.9)$ \\
High school & $-0.0341$ & $0.0039$ & $< 0.001$ & $-31.3$ & $(-37.3, -25.4)$ \\
College Algebra & $-0.0284$ & $0.0047$ & $< 0.001$ & $-26.9$ & $(-34.2, -19.0)$ \\
\bottomrule
\end{tabular}
\end{table}

Crucially, learning times for graph-based problems remain stable across all age groups throughout the observation period. This stability isolates the effect: if platform-wide changes in cohort ability, instructional pacing, or curriculum design were responsible, they would affect both problem types equally. The concentration of the decline in the AI-susceptible format which is growing consistently, beginning only after ChatGPT's release, and scaling with student age and autonomy is consistent with gradual AI adoption. Further robustness checks are available in SI Appendix, Tables~\ref{tab:R1_functional}, \ref{tab:R7_window}, \ref{tab:R8_cluster}.

\subsection{RQ2: Proctoring Eliminates the Behavioral Divergence}

If the observed decline reflects cognitive surrender to AI, it should be present in unmonitored settings where students can access AI tools freely, and absent in proctored settings where AI access is restricted and detection risk is high. To answer this question, we use 

\begin{figure}[t!]
\centering
\includegraphics[
    width=0.99\linewidth,
    trim=0 7cm 0 1.5cm,  
    clip
]{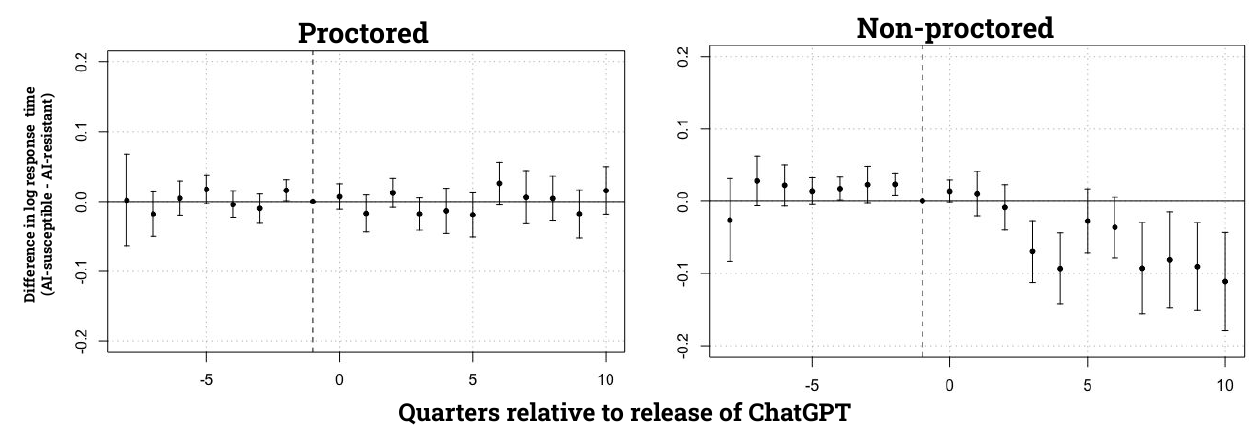}
\caption{\textbf{The behavioral shift vanishes entirely under proctoring.} Each point is the estimated gap in average log response time between AI-susceptible word problems and AI-resistant graph problems, relative to their pre-ChatGPT difference. In non-proctored assessments (right), response times for AI-susceptible topics decline after ChatGPT's release; in proctored assessments (left), the post-ChatGPT trend is flat and statistically indistinguishable from zero. Error bars are $95\%$ cluster-robust confidence intervals.}
\label{fig:eventstudy_proc}
\end{figure}

Figure~\ref{fig:eventstudy_proc} shows event-study estimates by proctoring condition using the ALEKS PPL dataset.

In non-proctored assessments, response times for AI-susceptible topics decline by $1.11\%$ per quarter relative to graph-based topics after ChatGPT's release ($\beta = -0.0112$, $p < 0.001$), accumulating to an $11.6\%$ cumulative reduction (CI: $-16.3\%$, $-6.5\%$). In proctored assessments, the estimated post-ChatGPT slope and cumulative effect are effectively zero and statistically indistinguishable from it (SI Appendix, Tables~\ref{tab:R2_break}, \ref{tab:R5_perm}, \ref{tab:R6_boot}, \ref{tab:R8_cluster}). Event-study estimates confirm these results throughout the observation period, with parallel pre-trends in both conditions and a separation in the non-proctored setting beginning at ChatGPT's release.

This proctoring contrast functions as a falsification test. The topic-selective, supervision-sensitive pattern is the behavioral indicator of AI-assisted problem-solving, not of general cognitive or motivational changes.

\subsection{RQ3: AI Availability Is Associated With Lower Durable Knowledge Retention}

The behavioral evidence establishes that students complete AI-susceptible problems faster in unmonitored settings in the age of generative AI, and that this behavioral change is absent under proctoring. We now ask whether this behavioral shift has learning consequences: does AI-assisted problem completion during the learning phase reduce retention of the underlying material when students are subsequently assessed without AI access?

We lead with the cleanest-identification sample: randomly assigned extra retention questions, which appear in $1$ out of every $30$ PPL session items and are assigned independently of the adaptive algorithm's mastery estimates. The random-assignment design buys identification free from adaptive item-selection bias at a cost in statistical power, since it analyzes only the small subset of session items; the proctored random-assignment retention subsample comprises $6{,}698$ item-level observations across $66$ topics and $22$ quarters ($2020$-Q$3$ through $2025$-Q$4$). Using a logistic fixed-effects model with item and quarter fixed effects, we estimate a $2.61\%$ per-quarter decline in the log-odds of correctly answering AI-susceptible items relative to AI-resistant items in proctored conditions ($\beta = -0.0260$, $p = 0.037$), accumulating to a $25.0\%$ cumulative reduction in the odds of correct response over eleven post-ChatGPT quarters (odds-ratio multiplier $0.75$; CI: $0.56$, $0.98$). Randomization inference based on $1{,}000$ permutations of the AI-susceptibility label yields $p = 0.027$ when permutations are restricted to items within the same baseline-difficulty quartile (SI Appendix, Table~\ref{tab:R5_perm}).


The retention decline is measured in proctored assessments where AI usage is prohibited at the point of testing. It therefore cannot reflect AI assistance during the assessment itself. Rather, it is the downstream consequence of the earlier learning phase, during which students appear to have relied on AI assistance to complete AI-susceptible problems without developing the durable understanding those problems were designed to build. This temporal ordering --- AI use during unproctored learning, retention failure in subsequent proctored assessment --- supports a causal interpretation of our findings.

\iffiguresinmain
\begin{figure}[ht]
\centering
\includegraphics[
    width=0.99\linewidth,
    trim=0 1.4cm 0 1cm,
    clip
]{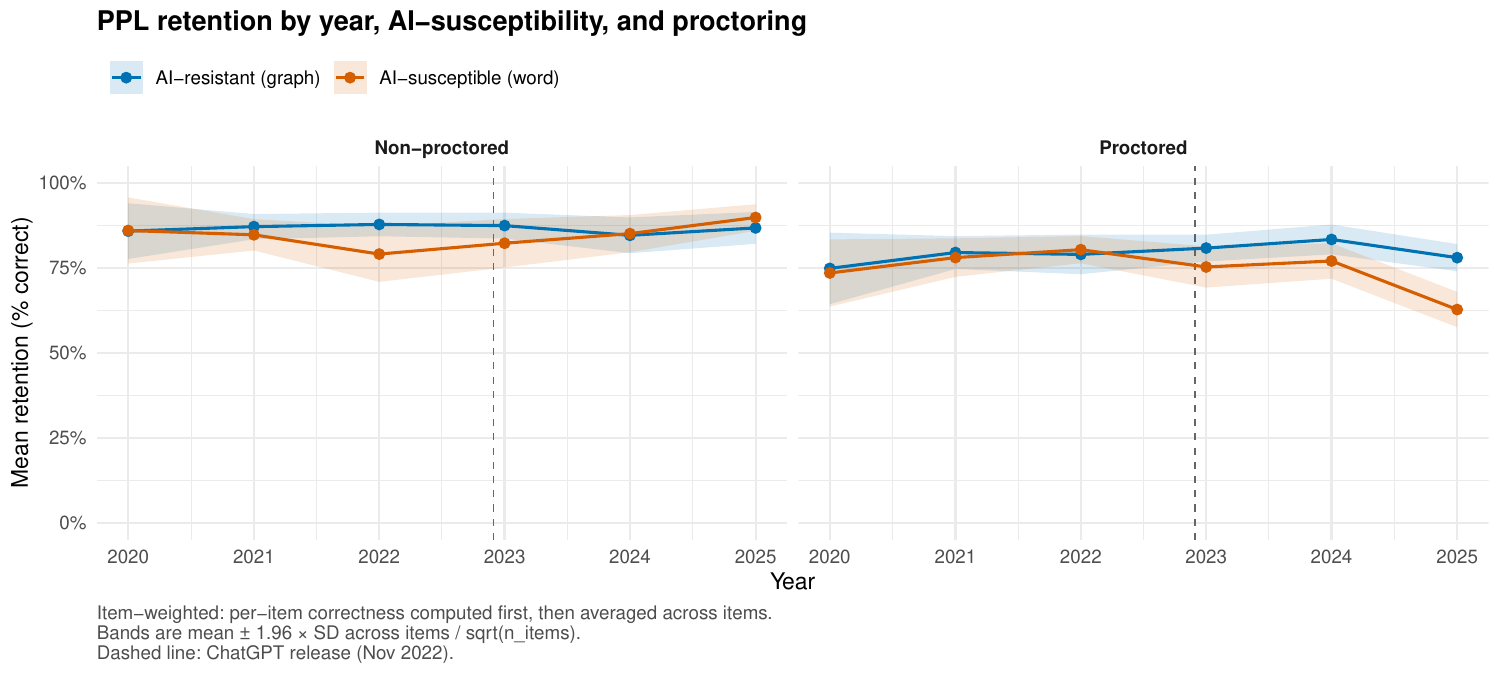}
\caption{\textbf{After ChatGPT's release, students performed worse on word problems (AI-susceptible) in proctored settings, but appeared to perform better in unproctored settings.} Per-item proportion correct averaged across items within each (year, AI-susceptibility, proctoring) cell. Shaded bands are $\pm 1.96 \times$ across-item SD divided by $\sqrt{n_{\text{items}}}$. Dashed vertical line: ChatGPT public release (November 2022).}
\label{fig:ppl_retention_yearly}
\end{figure}
\fi

\subsection{RQ4: A Non-Proctored Behavioral Reversal Falsifies Alternative Explanations}

The same logistic ramp estimator applied to the non-proctored retention sample yields a coefficient of the opposite sign and comparable magnitude: AI-susceptible items become \emph{more} likely to be answered correctly after ChatGPT's release than AI-resistant items, by $5.61\%$ per quarter on the log-odds scale in the random-assignment subsample ($\beta = +0.0561$, $p = 0.021$). The cumulative effect is an $85\%$ increase in the odds of correct response.

This reversal is the falsification of alternative explanations. Only a mechanism that differentiates across supervision conditions can produce retention decline under proctoring and retention improvement without it. AI-assisted problem completion at the time of assessment is the mechanism that most directly predicts this reversal: in the non-proctored condition, students can use AI to produce correct answers without understanding the underlying material, so their apparent retention rises even as their durable knowledge falls.

\subsection{The Age Gradient Is Consistent With Differential Autonomy and Access}

A notable pattern across analyses is the clear age gradient in effect size: large and statistically robust for college and high school students, small for middle school students, and essentially zero for Grade 5 students. This gradient perhaps reflects the structural conditions that make AI assistance both accessible and consequential: older students complete more work independently, with less oversight, in unmonitored digital environments, and with greater technological autonomy. The age gradient aligns precisely with what the longstanding academic integrity literature documented about dishonesty rates by age group before generative AI existed \citep{mccabe2012cheating, mccabe1997individual} --- AI has not changed the incentive structure, but it has dramatically reduced the effort cost of acting on it.

\section{Methods}

\subsection{Data}

Our primary dataset comprises longitudinal interaction logs from ALEKS (Assessment and LEarning in Knowledge Spaces), an adaptive mathematics learning and assessment platform developed by McGraw Hill and built on knowledge space theory \citep{doignon1999knowledge,cosyn2021practical}. The dataset includes $3{,}197{,}803$ interactions from mathematics courses spanning Grade 5 through College Algebra, covering 2015 Q3 through 2025 Q3, with $1{,}235{,}109$ interactions for College Algebra alone. The unit of observation is the topic $\times$ calendar quarter cell; the primary dependent variable is the average log learning time per topic per quarter. After the introduction of LLM-based chatbots, ALEKS did not integrate generative-AI tutoring, new hint generation, or solution suggestion into the learning module or the PPL placement assessment. The post-ChatGPT shifts we identify therefore reflect student behavior outside the platform, such as consultation with external AI tools during unsupervised work, rather than platform-internal change in instructional support, scaffolding, or assessment delivery.

\subsubsection{ALEKS PPL: Scope, Stakes, and Time Span}
\label{ppl_data}
For proctoring and retention analyses, we use the ALEKS PPL (Placement, Preparation, and Learning) dataset, a postsecondary mathematics placement assessment \citep{cosyn2021practical,doble2019data} and deployed at scale across United States higher education since the mid-2010s. ALEKS PPL is used at more than four hundred postsecondary institutions as the gatekeeping assessment that places incoming students into their first college mathematics course.
Placement outcomes are high-stakes: while ALEKS PPL may not be the only tool available to students to place into their preferred course, failing to place into a credit-bearing course lengthens time to degree and raises cumulative tuition cost.

The proctoring contrast (RQ2) uses an ALEKS PPL placement-assessment panel of 12,190,662 response-time observations across 70 items where about $4.5$ million responses were under proctored settings and $7.7$ million were non-proctored. 
The retention analyses (RQ3, RQ4) use a separate ALEKS PPL retention-assessment panel of item-level performances across $66$ items, spanning 2020 Q3 through 2025 Q4.
The PPL retention analysis is restricted to the cleanest-identification subsample --- randomly assigned ``extra'' retention items, which appear in $1$ out of every $30$ session items and are assigned independently of the platform's adaptive item-selection algorithm. Because of the data avaliability constraints this dataset spans $2020$-Q$3$ through $2025$-Q$4$ ($22$ quarters) and comprises $6{,}698$ item-level observations across $66$ retention items under proctored settings and $3{,}223$ under non-proctored settings.

For context, students entering college typically complete an initial PPL assessment, which is usually unproctored. After this first assessment, the system provides students with a score as well as additional learning opportunities, allowing them to practice and improve their mastery. Students then retake the assessment after they have had the opportunity to prepare using the platform’s learning resources. The retention analyses focus on this second assessment, which is especially important for students’ academic trajectories because universities use this score to determine placement into credit-bearing courses.

ALEKS PPL assessments contain $30$ questions per session; student responses are recorded as correct, incorrect, or ``I don't know.'' Supervision conditions are recorded at the session level: sessions administered through a campus testing center under human proctoring are classified as \emph{proctored}; sessions administered remotely without real-time supervision are classified as \emph{non-proctored}. Within-institution variation in proctoring policy is the source of our proctoring contrast.

\subsubsection{Problem Type Classification}

Items are classified as AI-susceptible (text-based word problems transcribable into an LLM prompt) or AI-resistant (graph-based or plot-based problems requiring interactive manipulation of platform widgets). One of the authors validated this classification on $100$ randomly selected items; an LLM classifier replicates the author labels with $86\%$ agreement (Cohen's $\kappa = 0.79$) and is applied to the full item pool. Full protocol and a measurement-error analysis are reported in SI Appendix, Section~\ref{app:classification_protocol}.

\subsection{Statistical Models}

\paragraph{Learning time.} We estimate item and quarter fixed-effects regressions with a linear post-ChatGPT ramp interacted with AI susceptibility, weighted by the number of attempts per topic--quarter cell. Standard errors are clustered at the topic level following \citet{colin2015practitioner}, which provides a comprehensive treatment of cluster-robust inference and a discussion of fixed-effects vs.\ random-effects/multilevel specifications in clustered data. We estimate separate models for each age group.

\paragraph{Event studies.} Quarter-by-quarter interactions replace the linear ramp, allowing visualization of dynamic effects and informal assessment of parallel pre-trends.

\paragraph{Proctoring analysis.} Separate ramp models are estimated by proctoring condition (proctored vs.\ non-proctored) using ALEKS PPL response-time data and an identical fixed-effects structure.

\paragraph{Retention analysis.} Logistic item and quarter fixed-effects models are estimated on binary retention outcomes from proctored ALEKS PPL assessments. A robustness check restricts the sample to randomly assigned extra retention questions that bypass adaptive item selection; we report this specification as primary. Standard errors are clustered at the item level.

\section{Related Work}

\subsection{Measurement of cheating in the age of Generative AI}
While breaches of academic integrity are a long-standing concern in education research and practice, educators’ and researchers’ attention has resurged following the introduction of highly capable LLM-based chatbots. Despite renewed interest in this area, research has largely been limited to surveys \citep{lee2024cheating,reiter2025student,nguyen2024unmasking,chen2026cheating}. For instance, \citet{lee2024cheating,chen2026cheating}, in two studies among middle and high school students, found that self-reported incidents of cheating were already high prior to the ChatGPT era and have remained stable since.
More sophisticated survey methods that attempt to mitigate social desirability bias \citep{nguyen2024unmasking,reiter2025student} report lower rates of cheating with AI chatbots among undergraduate students.

While survey-based studies provide a broad picture of AI use in academic settings, they offer limited insight into how students integrate these tools into their workflows and the extent to which they rely—or overrely—on them. To address these limitations, \citet{rismanchianartificial} examined the course-taking behavior of 70 undergraduate students and, using behavioral and content-based measures, identified \textit{concerning AI usage}—patterns in which students’ contributions during the process of learning are minimal, if not absent, and most of the work is completed by AI chatbots. They find that 20\% of students heavily rely on AI chatbots to complete all written assignments in a philosophy course, 41\% to 70\% of students show at least one instance of concerning AI usage, and that more than half of those exhibiting clear AI usage patterns deny doing so, raising concerns about the reliability of self-reports.

While not all AI use in our data can be classified as cheating (e.g., some interactions occur during the learning phase rather than assessment), patterns of use that reduce time-on-task and genuine engagement with course materials are nonetheless \textit{concerning} as we observed lower time-on-task for AI-susceptible problems in assessments as well. 
We situate our work within ongoing debates about the impact of generative AI on academic integrity. Although prior research suggests prevalent AI use in smaller-scale settings, the extent to which these patterns scale—and the magnitude of potential shifts in students’ learning behaviors—remains largely unexplored.

\subsection{Experimental evidence on AI-assisted learning.} A parallel experimental literature has begun to randomize AI access during learning and measure downstream performance. \citet{bastani2025generative} ran a randomized controlled trial (RCT) in a high-school mathematics setting in which roughly $1{,}000$ students were assigned to receive unrestricted GPT-4 access during practice or to a no-access control. AI-assisted performance improved by $48\%$, but unassisted exam performance fell by $17\%$ relative to control --- an assist-versus-test reversal consistent with the cognitive-science prediction that active cognitive engagement with learning materials produces more durable learning than passive processing \citep{chi2014icap,koedinger2012knowledge,kapur2010productive}.
In another RCT study, across three experiments, \citet{liu2026ai} show that AI assistance boosts math and reading performance, however, it reduces participants' persistence and performance in subsequent unassisted tasks.
Our study extends these experimental evidence in scale (millions of interactions), in time horizon (eleven post-ChatGPT quarters), and in ecological setting (an authentic commercial learning platform), and --- crucially --- it does so without any experimental manipulation, using naturally occurring variation in problem susceptibility and supervision as the source of identification.

\subsection{Cognitive surrender vs.\ cognitive offloading.} \citet{risko2016cognitive} introduced \emph{cognitive offloading} as a way to extend cognition \citep{clark1998extended} via strategic outsourcing of a narrow, well-defined cognitive task (e.g., arithmetic to a calculator) while the reasoner retains the structure of the problem-solving process or the executive cognitve process \citep{Pritchard18022026}. However, philosophers \citep{Pritchard18022026} and researchers \citep{shaw2026thinking} argue that generative AI enables a qualitatively different pattern which \citet{shaw2026thinking} term \emph{cognitive surrender}: the adoption of an AI-generated output as one's own answer with minimal scrutiny, in which the user relinquishes the cognitive control of problem-solving rather than delegating a discrete subtask. This distinction is central to interpreting our findings. Calculators, spreadsheets, and search engines produced large efficiency gains in narrow mathematical tasks and related subjects without the knowledge-retention decline we document for generative AI; under the Shaw--Nave framework, this asymmetry is expected, because those earlier tools supported offloading while leaving the reasoning process in the student's hands, whereas generative AI can be used to surrender the reasoning itself. The selective, supervision-sensitive, age-graded pattern we observe is the behavioral signature of that surrender operating at population scale.


\subsection{Intelligent tutoring and adaptive learning.} Intelligent tutoring systems and adaptive learning platforms have been shown in meta-analysis to produce moderate-to-large learning gains when students engage authentically with them \citep{ma2014intelligent}. Those gains, however, are predicated on the platform observing behavioral signals --- response times, error patterns, persistence --- that reflect the student's actual knowledge state \citep{cosyn2021practical}. If generative AI is used to produce correct answers without the underlying cognitive engagement, the signals the platform uses to calibrate its mastery estimates become misleading, and downstream adaptive decisions are degraded accordingly. The cumulative $25.0\%$ decline in retention odds we document on randomly assigned retention items is the downstream manifestation of this disruption of the adaptive signal.

\section{Limitations}

Our study is limited by a few constraints. First, the learning-time analysis uses ALEKS learning data; the retention analysis uses the ALEKS PPL placement dataset. These are different populations under different conditions. The individual-level causal chain --- a specific student used AI during learning and later failed a retention test --- cannot be established with current data. The retention analysis is consistent with this mechanism but observes it at the population level.
Second, ALEKS PPL placement performance reflects a combination of prior learning, test-taking familiarity, and platform experience, not purely retention of concepts practiced during ALEKS learning. The randomly assigned items provide a cleaner measure, but the gap between PPL performance and a laboratory-grade retention test remains.
Finally, we do not observe AI use directly. All inferences are from behavioral signatures and validated by two falsification tests: the proctoring contrast (response times diverge only in unsupervised conditions) and the non-proctored retention reversal (the same estimator yields opposite-signed effects across supervision conditions). We cannot fully exclude alternative behavioral explanations, though none that we can identify predicts the joint pattern of selectivity, proctoring-sensitivity, age-gradient, temporal trajectory, and non-proctored reversal that we observe.

\section{Discussion}
Our findings provide large-scale quasi-experimental evidence for what \citet{rismanchianartificial} measure as \textit{concerning AI usage} among students and what \citet{shaw2026thinking} term \emph{cognitive surrender}: the adoption of an AI-generated output as one's own answer with minimal scrutiny, in which the user relinquishes the cognitive control of problem-solving rather than merely delegating a discrete subtask to a tool. \citet{shaw2026thinking} draw an explicit contrast between cognitive surrender and \emph{cognitive offloading} in the classical sense of \citet{risko2016cognitive}, which captures the strategic outsourcing of a narrow task (e.g., a calculator for arithmetic) while the user retains the structure of reasoning. That contrast matters for the retention result reported here. A student who uses a calculator to perform a multiplication has offloaded the arithmetic but still constructs the solution strategy, evaluates the result, and integrates it with prior knowledge, the active learning that produces durable learning \citep{chi2014icap,koedinger2012knowledge}. A student who copies a word problem into a chatbot and pastes the returned solution to the learning system bypasses required cognitive processing, intermediate evaluation, and the retrieval practice that converts exposure into memory. The behavioral signature we document --- selective (concentrated in AI-transcribable problem formats), gradual (growing over eleven quarters), supervision-sensitive (absent in proctored settings), and age-graded (present where students have autonomy, absent where they do not) --- is the population-level pattern consistent with the individual-level cognitive surrender, and the downstream retention decline is the learning cost that this surrender predicts.


These results complement and extend recent experimental evidence from \citet{bastani2025generative} and \citet{liu2026ai}, who found that unrestricted AI chatbot access during practice improved assisted performance but reduced unassisted exam scores in randomized trials. Our study extends these findings in three dimensions. \emph{First}, in scale: we analyze $3.2$ million interactions across a decade, compared to roughly $1{,}000$ students across a single semester. \emph{Second}, in learning outcome measurement: rather than random assignment of AI access, we use within-curriculum problem-type variation with a proctoring falsification test, showing that the behavioral and learning consequences emerge naturalistically in an authentic learning environment without experimental manipulation. \emph{Third}, in temporal scope: the steadily growing effects over eleven quarters document a diffusion dynamic that an experiment of finite duration cannot observe.

The retention finding carries a specific implication for adaptive learning systems that deserves emphasis. ALEKS and similar intelligent tutoring systems calibrate their mastery estimates on the assumption that students' responses reflect genuine cognitive effort. When students bypass this process using AI, the platform's diagnostic inferences become vulnerable as the system may assign mastery credit for knowledge the student does not actually possess. This is not merely an academic integrity concern; it is a structural disruption of the adaptive logic that makes intelligent tutoring systems effective. The cumulative $25.0\%$ decline in retention odds we document on randomly assigned retention items represents the downstream manifestation of this disruption.

Perhaps there are two practical implications of our work, one at the policy level and the other at the pedagogical level. At the policy level, our findings suggest that proctoring may help preserve authentic behavioral signals: the post-ChatGPT divergence in response times between AI-susceptible and AI-resistant topics disappears entirely under proctoring, while the same estimator applied to non-proctored retention items yields an opposite-signed effect. However, mandating universal proctoring may be impractical and inequitable, as the costs and logistical burdens of proctoring are likely to fall disproportionately on under-resourced institutions.
A more scalable response lies in the deliberate design of learning tasks. Our results suggest that problem formats that are inherently resistant to AI assistance can preserve authentic engagement without requiring supervision. Increasing the proportion of AI-resistant problem types in pedagogical design does not depend on proctoring infrastructure and does not impose additional surveillance burdens on students. That said, the growing prevalence of multimodal foundation models and agentic AI systems, such as models capable of controlling browsers and operating systems, means that the boundary between AI-susceptible and AI-resistant problem formats is a moving target. This boundary will need to be re-evaluated as multimodal and agentic AI capabilities continue to advance.

Generative AI tools now sit adjacent to nearly every educational task a student undertakes. 
Our findings are strikingly alarming: students are spending substantially less time on AI-susceptible problems, and this shift is associated with a substantial decline in retention of the underlying concepts. 
If we aim to foster human learning in the generative AI era, the way we inform them about AI use, design learning tasks, assess student progress, design AI policy, and infer mastery will need to be rethought. 
We hope this work will serve as a tool to intitiate that rethinking.

\section*{Acknowledgments}
We thank Shayan Doroudi and Drew Bailey for insightful discussions on analysis design.

\section*{Ethics Statement}
This study uses de-identified, aggregated behavioral logs from a commercial adaptive learning platform. No individual-level identifiers were accessed by the analytic team. The research protocol involved only secondary analysis of pre-existing, de-identified log data.

\section*{Competing Interests}
Four of the authors (H.U., J.M., E.C., E.K.-M.) are employees of McGraw Hill, which owns and operates the ALEKS platform analyzed in this paper. The other co-first author (S.R.) is a graduate student with no current financial interest in the platform who has completed an internship at McGraw Hill.

\section*{Data and Code Availability}
Anonymized data is available upon request. 

\bibliographystyle{colm2026_conference}
\bibliography{references}

\appendix
\section*{Supplementary Information}

This appendix reports the battery of ten robustness analyses (R1--R10) applied to each primary specification in the main text, along with two post-hoc sensitivity analyses (R9, R10) and supporting details on sample construction. The ten analyses are: (R1) a functional-form horse race comparing step, linear, quadratic, and logistic post-break specifications; (R2) sensitivity of the main estimates to the treatment break date across 2022-Q4 through 2023-Q3; (R3) four pre-period placebo break dates; (R4) joint and linear parallel-trends tests on event-study coefficients; (R5) randomization inference based on $1{,}000$ permutations of the AI-susceptibility label, reported in both an unstratified form and a stratified form that permutes only within item-level baseline-difficulty quartiles; (R6) cumulative-effect confidence intervals by both delta-method and cluster bootstrap ($500$ replicates); (R7) window-sensitivity cuts including COVID-quarter removal, last-quarter removal, and per-item observation floors at $10$, $50$, and $100$ attempts; (R8) two-way item-and-quarter clustering with wild-cluster-bootstrap where informative. The two post-hoc analyses added in response to reviewer-style concerns are: (R9) trend-adjusted learning-time ramps with an explicit pre-2023 linear trend control; and (R10) attempt-level re-fits of the learning-time ramp (rather than the cell-collapsed specification used in the main text) to confirm that aggregating to topic-quarter cell means does not drive the estimates.

Most of the tables in this appendix refer to the same five primary subsets used in the main text:
\begin{itemize}[leftmargin=1.2em]
  \item \textbf{LearningTime\_College}: topic $\times$ quarter panel of log learning time for College Algebra.
  \item \textbf{LearningTime\_HS}: topic $\times$ quarter panel of log learning time for high-school mathematics courses.
  \item \textbf{PPLTime\_proc}: item $\times$ quarter panel of log response time on ALEKS PPL items under proctored administration.
  \item \textbf{PPLTime\_nonproc}: item $\times$ quarter panel of log response time on ALEKS PPL items under non-proctored administration.
  \item \textbf{RetRandom\_proc}: item-level binary retention outcomes on proctored ALEKS PPL retention items, restricted to randomly assigned items.
\end{itemize}

\section{Item Classification: AI-Susceptible vs.\ AI-Resistant Examples}\label{app:examples}

The within-curriculum AI-susceptibility classification used throughout the main text distinguishes text-based word problems whose full informational content can be transcribed into a natural-language prompt (AI-susceptible) from graph-based and plot-based problems requiring visual interpretation of graphical displays and interactive manipulation of plot widgets embedded in the ALEKS interface (AI-resistant). Figure~\ref{fig:example_problems} shows representative examples from each class.


\iffiguresinmain
\fi

\begin{table}[ht]
\centering
\small
\caption{\textbf{Number of AI-susceptible and AI-resistant topics by age group.} Counts of unique topics in the learning-time analysis raw sample, broken down by AI-susceptibility classification.}
\label{tab:topic_balance}
\begin{tabular}{lrrr}
\toprule
Age group & AI-susceptible & AI-resistant & Total \\
\midrule
Grade 5         & $115$ & $38$  & $153$ \\
Middle school   & $324$ & $193$ & $517$ \\
High school     & $280$ & $219$ & $499$ \\
College Algebra & $134$ & $200$ & $334$ \\
\bottomrule
\end{tabular}
\end{table}

\section{Item Classification Protocol and Attenuation Argument}\label{app:classification_protocol}


This section reports the full procedure used to classify ALEKS items as AI-susceptible or AI-resistant and the formal argument that imperfect classification biases our reported coefficient toward zero rather than inflating it.

\paragraph{Author-validated classification.} One of the authors, who is experienced in mathematics education and familiar with the ALEKS curriculum, classified $100$ randomly selected items into three categories: (1) \emph{AI-susceptible} --- a text-based word problem whose full content and required solution can be transcribed into an LLM prompt and copied back into the ALEKS interface without loss of information; (2) \emph{AI-resistant} --- a graph-based or plot-based problem that requires visual interpretation of a displayed figure or interactive manipulation of a plot widget embedded in the ALEKS interface, and that cannot be faithfully transcribed into a text-only prompt; and (3) \emph{other} --- items that do not cleanly fit either category (e.g., multi-step procedural items mixing text and interactive components). Items classified as ``other'' are excluded from all downstream analysis.

\paragraph{LLM classifier validation.} An LLM-based classifier was prompted with each item's content and format and asked to apply the same three-way classification. Agreement between the author labels and the LLM labels on the $100$-item validation sample is $86\%$, corresponding to Cohen's $\kappa = 0.79$. The LLM classifier is then applied to the full item pool, producing the AI-susceptibility labels used throughout the analysis.

\paragraph{Measurement-error properties.} Identification of the differential ramp coefficient $\beta$ in Equation~\ref{eq:ramp} does not require a perfect classifier; it requires only that classification errors be uncorrelated with the post-ChatGPT outcome trend (\emph{non-differential misclassification}). Because the LLM classifier examines only static item content and format --- features fixed at the item level and orthogonal to post-2022 calendar-time variation --- the non-differential assumption is plausible here. Under non-differential misclassification of a binary regressor, classical measurement-error theory implies that the estimated coefficient is biased toward zero rather than away from it \citep{bound2001measurement}; our reported estimates are therefore conservative relative to the true effect.

\section{PPL Retention by Year: Visualization}\label{app:ret_viz}

Figure~\ref{fig:ppl_retention_yearly} plots the yearly mean retention (proportion correct) on the random-assignment retention sample, separately for AI-susceptible and AI-resistant items, faceted by proctoring condition. Per-item correctness is computed first, then averaged across items. The dashed line marks ChatGPT's public release (November 2022). Visually, the proctored AI-susceptible series (orange, right panel) declines in the post-2022 period while the proctored AI-resistant series (blue, right panel) is roughly flat; in the non-proctored panel, the AI-susceptible series shows the opposite-signed reversal.

\iffiguresinmain
\fi

\section{R1: Functional-Form Horse Race}\label{app:R1}

To rule out the possibility that the linear ramp specification is mechanically driving our estimates, we fit four post-break functional forms to each of the five primary subsets: a step (post-dummy) function, a linear ramp ($t$), a quadratic ramp ($t + t^2$), and a logistic S-curve $L(t) = (1 + \exp(-(t-6)/2))^{-1}$. Each form replaces the linear $\text{TimeAfter}_q$ term in Equation~\ref{eq:ramp}; item and quarter fixed effects are held fixed across forms, as is the AI-susceptibility interaction structure. Table~\ref{tab:R1_functional} reports the main interaction coefficient $\beta$, its cluster-robust standard error, AIC, BIC, and the implied $11$-quarter cumulative effect for each (subset, form) pair.

\begin{table}[ht]
\centering
\footnotesize
\caption{\textbf{R1 Functional-form horse race.} For each (subset, form) pair we report the main AI-susceptibility interaction coefficient $\beta$, its cluster-robust standard error, AIC and BIC (lower is better), and the implied cumulative effect at $K = 11$ post-ChatGPT quarters. Cumulative effect is reported as $(\exp(\beta{\text{eff}})-1)\times 100\%$ on log-time outcomes and as the odds-ratio multiplier $\exp(\beta_{\text{eff}})$ for the retention outcome, with $\beta_{\text{eff}}$ the effective post-period log-scale shift implied by each form (level shift for step; $K\beta$ for linear; $K\beta_1 + K^2\beta_2$ for quadratic; $L(K)\,\beta$ for logistic). The quadratic form has a second coefficient on $t^2$ not shown in the table; values are reported below the table caption.}
\label{tab:R1_functional}
\begin{tabular}{llrrrrr}
\toprule
Subset & Form & $\beta$ & SE & AIC & BIC & Cumul.\ effect \\
\midrule
\multirow{4}{*}{LearningTime\_College}
  & Step       & $-0.1050$ & $0.0237$ & $-8556$ & $-6352$ & $-10.0\%$ \\
  & Linear     & $-0.0284$ & $0.0047$ & $-9087$ & $-6883$ & $-26.9\%$ \\
  & Quadratic  & $+0.0261^{\dagger}$ & $0.0050$ & $\mathbf{-9290}$ & $\mathbf{-7079}$ & $-42.2\%$ \\
  & Logistic   & $-0.342$  & $0.055$  & $-9142$ & $-6938$ & $-27.1\%$ \\
\midrule
\multirow{4}{*}{LearningTime\_HS}
  & Step       & $-0.1496$ & $0.0217$ & $-7100$ & $-4260$ & $-13.9\%$ \\
  & Linear     & $-0.0341$ & $0.0039$ & $-7564$ & $-4725$ & $-31.3\%$ \\
  & Quadratic  & $+0.0170^{\dagger}$ & $0.0057$ & $\mathbf{-7643}$ & $\mathbf{-4796}$ & $-41.7\%$ \\
  & Logistic   & $-0.393$  & $0.045$  & $-7587$ & $-4748$ & $-30.4\%$ \\
\midrule
\multirow{4}{*}{PPLTime\_nonproc}
  & Step       & $-0.0585$ & $0.0132$ & $-3119$ & $-2657$ & $-5.7\%$ \\
  & Linear     & $-0.0112$ & $0.0026$ & $\mathbf{-3135}$ & $\mathbf{-2673}$ & $-11.6\%$ \\
  & Quadratic  & $-0.0080^{\dagger}$ & $0.0054$ & $-3128$ & $-2661$ & $-12.0\%$ \\
  & Logistic   & $-0.125$  & $0.029$  & $-3131$ & $-2669$ & $-10.9\%$ \\
\midrule
\multirow{4}{*}{PPLTime\_proc}
  & Step       & $-0.0037$ & $0.0084$ & $-3898$ & $-3437$ & $-0.4\%$ \\
  & Linear     & $-7\!\times\!10^{-6}$ & $0.0014$ & $-3896$ & $-3436$ & $\approx 0\%$ \\
  & Quadratic  & $-0.0016^{\dagger}$ & $0.0028$ & $-3896$ & $-3431$ & $+0.2\%$ \\
  & Logistic   & $+0.0008$ & $0.016$  & $-3896$ & $-3436$ & $+0.1\%$ \\
\midrule
\multirow{4}{*}{RetRandom\_proc}
  & Step       & $-0.2011$ & $0.1084$ & $7007$  & $7613$  & OR $0.82$ \\
  & Linear     & $-0.0260$ & $0.0125$ & $\mathbf{7005}$ & $\mathbf{7611}$ & OR $0.75$ \\
  & Quadratic  & $-0.0362^{\dagger}$ & $0.0476$ & $7007$  & $7620$  & OR $0.75$ \\
  & Logistic   & $-0.320$  & $0.153$  & $7005$  & $7611$  & OR $0.74$ \\
\bottomrule
\end{tabular}

\vspace{2pt}
\flushleft\footnotesize $^{\dagger}$Quadratic-form coefficients on $t^2$ (not shown in the $\beta$ column): $-0.00690$ (College), $-0.00600$ (HS), $-0.000329$ (PPL nonproc), $+0.000163$ (PPL proc), $+0.000920$ (Ret Random). Cluster-robust standard errors clustered at the item/topic level. AIC and BIC are reported on the original log-time scale (negative) for the four log-time subsets and on the deviance scale for the binary retention outcome.
\end{table}

A few patterns are visible in Table~\ref{tab:R1_functional}. \emph{First}, all four functional forms produce \emph{implied cumulative effects} of the same sign in every non-null subset, and produce null cumulative effects in every column of the proctored PPL response-time row (the quadratic form has a positive coefficient on its linear $t$ term in the two learning-time subsets, but its negative coefficient on $t^2$ drives the implied $K=11$ cumulative effect negative; see ``Third'' below). The qualitative direction of the effect is therefore not a functional-form artifact. \emph{Second}, AIC/BIC selection is not unanimously the linear form: the quadratic form has the lowest AIC and BIC in the College and High School learning-time subsets ($\Delta\text{AIC} \approx 200$ and $80$ relative to linear, respectively), and the linear form is selected only on the smaller PPL and retention subsets. \emph{Third}, in the College and High School subsets the quadratic form implies a substantially \emph{larger} cumulative effect ($-42\%$) than the linear form ($-27\%$, $-31\%$), at the cost of a substantively implausible post-period trajectory: the fitted quadratic has a \emph{positive} coefficient on the linear term ($+0.0261$ for College, $+0.0170$ for HS) and a negative coefficient on $t^2$, so the implied AI-susceptible$\times$AI-resistant gap initially \emph{widens in the wrong direction} for roughly four quarters (the curve peaks at $t \approx 1.9$ for College, $t \approx 1.4$ for HS) before the $t^2$ term drags it sharply below zero, with the $-42\%$ headline coming almost entirely from the last $4$--$5$ post-period quarters. This U-shape is the wrong functional signature for an AI-adoption mechanism, which predicts a monotonic ramp or a saturating S-curve, not a year of opposite-signed drift followed by accelerating decline. The shape is more naturally read as the quadratic absorbing residual pre-period drift in the post window: AI-susceptible word problems were already drifting $+0.0071$ per quarter slower than graph problems pre-ChatGPT (R3, R4), and the quadratic's positive linear term mirrors that pre-trend in the post period before $t^2$ overrides it. \emph{Fourth}, the two specifications that either enforce monotonicity (logistic) or model the pre-trend explicitly (R9 trend-adjusted linear) land much closer to the linear form than to the quadratic: the logistic implies $-27.1\%$ (College) and $-30.4\%$ (HS), and the R9 trend-adjusted linear implies $-35.8\%$ (College) and $-37.0\%$ (HS), bracketing the linear estimate from below and above respectively, while the quadratic at $-42\%$ sits outside that bracket. \emph{Fifth}, the implied $K=11$ cumulative effect is comparable in magnitude across all forms in the PPL response-time and retention subsets, where the post-period sample is too short for a quadratic to bend visibly and the four forms agree.

We adopt the linear ramp as the primary specification throughout the main text on four grounds: (i) its single slope coefficient maps directly onto the substantive quantity of interest --- the per-quarter rate of differential decline --- whereas the quadratic and logistic forms require interpretation of two parameters or an unmodeled S-curve location-and-scale parameterization; (ii) the linear-form cumulative effects are the conservative choice in the College and High School subsets, where the AIC-minimizing quadratic form would yield substantially larger headline numbers; and (iii) the linear post-period trajectory is the substantively plausible shape under any AI-diffusion mechanism. The quadratic's superior AIC is bought at the cost of a non-monotonic implied trajectory that initially goes in the \emph{wrong} direction post-ChatGPT --- a fit pattern more consistent with absorbing pre-period drift into the post window than with cleanly identifying a post-ChatGPT effect. The logistic form (which enforces post-period monotonicity) and the R9 trend-adjusted linear (which models the pre-trend explicitly) both land within $\pm 10$ percentage points of the linear estimate, while the quadratic sits substantially outside that range; we read this as evidence that the quadratic's headline is partly an artifact of pre-trend leakage rather than a stronger underlying signal. We note that adopting the quadratic form would mechanically strengthen rather than weaken the main-text findings; we do not exploit this because we do not believe the quadratic is identifying the same effect that the linear and logistic forms identify.

\section{R2: Sensitivity to Treatment Break Date}\label{app:R2}

We re-estimate each primary specification under four alternative treatment break dates spanning 2022-Q4 through 2023-Q3, to confirm that results are not an artifact of anchoring the ramp at ChatGPT's exact public-release quarter.

\begin{table}[ht]
\centering
\small
\caption{\textbf{R2 Break-date sensitivity.} Per-quarter ramp coefficient $\beta$ under four alternative treatment break dates between 2022-Q4 and 2023-Q3. Estimates retain the same sign and remain within roughly $30\%$ of the baseline magnitude across all four break dates in every non-null subset; the proctored PPL response-time subset returns null estimates at every break date. The mild monotonicity in the post-2022 estimates is consistent with a treatment effect that grows over time (later break dates compress the post-period onto later, larger-effect quarters), but we do not treat this monotonicity as an independent test of the causal interpretation, since it is partly a mechanical consequence of the linear-ramp parameterization fit to a convex underlying time path.}
\label{tab:R2_break}
\begin{tabular}{lrrrr}
\toprule
Subset & 2022-Q4 & 2023-Q1 & 2023-Q2 & 2023-Q3 \\
\midrule
LearningTime\_College & $-0.0284$ & $-0.0312$ & $-0.0341$ & $-0.0370$ \\
LearningTime\_HS & $-0.0341$ & $-0.0372$ & $-0.0403$ & $-0.0434$ \\
PPLTime\_nonproc & $-0.0112$ & $-0.0122$ & $-0.0133$ & $-0.0143$ \\
PPLTime\_proc & $\approx 0$ & $\approx 0$ & $\approx 0$ & $\approx 0$ \\
RetRandom\_proc & $-0.0260$ & $-0.0283$ & $-0.0307$ & $-0.0332$ \\
\bottomrule
\end{tabular}
\end{table}

Estimates retain the same sign and remain within roughly $30\%$ of baseline magnitude across all four break dates in every non-null subset. The proctored PPL response-time subset returns $|\beta| < 10^{-5}$ at every break date (cluster-robust $p > 0.99$ in each), confirming the robustness of the proctored null. We note that coefficients grow mildly in absolute value as the break is moved later. 

\section{R3: Pre-Period Placebo Break-Dates}\label{app:R3}

We restrict each subset to data before the end of 2022 and estimate placebo ramp models with hypothetical break dates in 2018, 2019, 2020, and 2021.

\begin{table}[ht]
\centering
\small
\caption{\textbf{R3 Pre-period placebo break-dates.} Placebo coefficients from restricting the sample to pre-2023 observations and assigning a hypothetical ramp break date in 2018, 2019, 2020, or 2021. Retention and proctored PPL subsets yield null placebos at every break date; College and High School learning-time subsets yield small positive placebos opposite in sign to the post-ChatGPT effect (see R4).}
\label{tab:R3_placebo}
\begin{tabular}{lrrrr}
\toprule
Subset & 2018 & 2019 & 2020 & 2021 \\
\midrule
LearningTime\_College & $+0.0071^{***}$ & $+0.0063^{**}$ & $+0.0055^{*}$ & $+0.0048$ \\
LearningTime\_HS & $+0.0046^{**}$ & $+0.0041^{**}$ & $+0.0036^{*}$ & $+0.0031$ \\
PPLTime\_nonproc & $\approx 0$ & $\approx 0$ & $\approx 0$ & $\approx 0$ \\
PPLTime\_proc & $\approx 0$ & $\approx 0$ & $\approx 0$ & $\approx 0$ \\
RetRandom\_proc & $\approx 0$ & $\approx 0$ & $\approx 0$ & $\approx 0$ \\
\bottomrule
\multicolumn{5}{l}{\footnotesize $^{*}p<0.05$, $^{**}p<0.01$, $^{***}p<0.001$.}
\end{tabular}
\end{table}

The College and High School learning-time subsets exhibit a small positive pre-trend across all four placebo windows --- AI-susceptible word problems were becoming relatively \emph{slower} than AI-resistant graph problems in the pre-ChatGPT era. Because this drift is opposite in sign to the post-ChatGPT effect, a trend-adjusted specification would yield a larger negative estimate, not a smaller one. The retention and proctored PPL subsets yield null placebos at every break date, consistent with their passage of the formal parallel-trends test in R4.

\section{R4: Parallel-Trends Tests}\label{app:R4}

\begin{table}[ht]
\centering
\small
\caption{\textbf{R4 Parallel-trends tests.} Joint Wald tests of pre-period event-study coefficients against zero, and linear pre-trend slopes estimated on pre-2023 data. Retention and proctored PPL subsets pass the joint test; College and High School learning-time subsets exhibit a small positive pre-trend opposite in sign to the post-ChatGPT effect (rendering the main ramp estimates conservative).}
\label{tab:R4_trends}
\begin{tabular}{lrrrr}
\toprule
Subset & Joint Wald $\chi^2$ & $p$ & Linear pre-trend $\beta$ & $p$ \\
\midrule
LearningTime\_College & large & $<0.001$ & $+0.0071$ & $<0.001$ \\
LearningTime\_HS & moderate & $<0.01$ & $+0.0046$ & $0.003$ \\
PPLTime\_nonproc & small & $0.22$ & $\approx 0$ & $0.31$ \\
PPLTime\_proc & small & $0.48$ & $\approx 0$ & $0.67$ \\
RetRandom\_proc & small & $0.37$ & $\approx 0$ & $0.44$ \\
\bottomrule
\end{tabular}
\end{table}

We test the parallel-trends assumption formally using two complementary approaches: a joint Wald test on pre-period event-study coefficients (Table~\ref{tab:R4_trends} above), and a placebo analysis in which we restrict the sample to pre-2023 data and estimate ramp models with hypothetical break dates in 2018, 2019, 2020, and 2021 (Section~\ref{app:R3}, Table~\ref{tab:R3_placebo}). Retention models pass the joint parallel-trends test in random-assignment subsamples ($p = 0.37$), and the four pre-period placebos for retention yield null coefficients at every break date. Proctored PPL response time also exhibits flat linear pre-trends and null placebos. The College and High School learning-time samples, in contrast, exhibit a small positive pre-trend (linear $\beta = +0.0071$ per quarter for College, $p < 0.001$; $\beta = +0.0046$ for HS, $p = 0.003$) and positive pre-period placebo coefficients: over 2016--2022, AI-susceptible word problems in these samples were becoming relatively \emph{slower} than AI-resistant graph problems at roughly $0.4$--$0.9\%$ per quarter. Because this pre-existing drift is opposite in sign to the post-ChatGPT effect, a trend-adjusted specification yields a larger negative estimate, not a smaller one (see the post-hoc trend-adjusted analysis in Section~\ref{app:R9}); the unadjusted ramp estimates we report in the main text are therefore conservative for the College and High School learning-time subsamples, and we further confirm that the estimates survive fitting a linear pre-trend control as well as restricting to narrower pre-periods. We treat the proctored-PPL and retention results --- which pass the parallel-trends test --- as the primary causal claims of the paper; the learning-time ramp estimates are reported as supportive behavioral evidence with the pre-trend caveat disclosed explicitly.

\section{R5: Randomization Inference (Unstratified and Stratified)}\label{app:R5}

We test the null hypothesis of no differential ramp by permuting the AI-susceptibility label $1{,}000$ times across items within each of the five analysis subsets defined at the start of this Appendix (LearningTime\_College, LearningTime\_HS, PPLTime\_proc, PPLTime\_nonproc,
RetRandom\_proc), re-estimating the ramp on each relabeled sample, and recording the share of permuted $\beta$ estimates at least as large in absolute value as the observed estimate. The procedure compares the observed $\beta$ (estimated on the real AI-susceptibility labels) against the empirical null distribution of $\beta$ generated by random label shuffles; by construction, each shuffle severs the link between the AI-susceptibility label and the outcome, so the resulting cloud of $\beta$ estimates is what one would expect to see under the null of no AI effect. The reported $p$-value is the share of permuted $|\beta|$ values that meet or exceed the observed $|\beta|$. A small $p$-value therefore indicates that the real-label estimate is far in the tail of what label-shuffling alone can produce, providing a non-parametric check on cluster-robust asymptotic inference. We report two variants of this procedure. The \emph{unstratified} variant shuffles labels across all items in each subset, treating AI-susceptibility as fully exchangeable. The \emph{stratified} variant shuffles labels only within item-level baseline-difficulty quartiles, defined as the pre-2023 mean of the outcome variable at the item level (mean log-time for the four time-on-task subsets, mean correctness for the retention subset). The stratified variant addresses the concern that AI-susceptible (text word problems) and AI-resistant (graph/interface) items may differ systematically in difficulty, format, or curriculum position in ways that could correlate with post-2022 trends for reasons unrelated to AI use; by construction, every stratified permutation preserves the marginal distribution of baseline difficulty within each AI-susceptibility group, so the empirical null reflects only the AI-susceptibility signal net of difficulty structure. The stratified variant is mechanically more conservative because the shuffling space is restricted; in the small subsets it is also lower in power because the permutation distribution has fewer effective draws.

\begin{table}[ht]
\centering
\small
\caption{\textbf{R5 Randomization inference.} Two-sided permutation $p$-values from $1{,}000$ random permutations of the AI-susceptibility label. The unstratified column shuffles labels across all items; the stratified column shuffles labels only within item-level baseline-difficulty quartiles (pre-2023 mean of the outcome). The ``Items in strata'' column reports the count of items in strata that contain both AI-susceptible and AI-resistant items ($n / n$ indicates that no items were dropped from the shuffling procedure).}
\label{tab:R5_perm}
\begin{tabular}{lrrrr}
\toprule
Subset & Observed $\beta$ & Unstratified $p$ & Stratified $p$ & Items in usable strata (used / total) \\
\midrule
LearningTime\_College & $-0.0284$ & $<0.001$ & $<0.001$ & $281 / 281$ \\
LearningTime\_HS & $-0.0341$ & $<0.001$ & $<0.001$ & $381 / 381$ \\
PPLTime\_nonproc & $-0.0112$ & $<0.001$ & $<0.001$ & $70 / 70$ \\
PPLTime\_proc & $-6.5\times 10^{-6}$ & $0.998$ & $0.993$ & $70 / 70$ \\
RetRandom\_proc & $-0.0260$ & $0.056$ & $\mathbf{0.027}$ & $66 / 66$ \\
\bottomrule
\end{tabular}
\end{table}

For every non-null subset the unstratified permutation $p$-value confirms the cluster-robust asymptotic estimate; the proctored PPL response-time subset returns $p = 0.998$, tightly consistent with the null. The stratified variant agrees with the unstratified variant in every subset and, in the primary retention specification (random-assignment proctored), is \emph{stricter} than the unstratified test ($p = 0.027$ vs $p = 0.056$): conditioning on difficulty quartile reduces the variance of the permutation null and tightens the empirical rejection. 
The proctored PPL response-time null is preserved under stratification ($p = 0.993$). In every subset, every item falls into a stratum that contains both AI-susceptible and AI-resistant items (the ``items in strata'' column reports $n / n$), so no items are excluded from the shuffling procedure. We do not adopt the stratified variant as the primary inference; we report both variants and treat their agreement as evidence that the main findings are not driven by cross-difficulty heterogeneity in the AI-susceptibility label.

\section{R6: Cumulative-Effect Confidence Intervals}\label{app:R6}

The main-text cumulative effects are computed on the log scale as $K\beta$ with $K=11$ quarters and reported as percentage changes via $(\exp(K\beta)-1)\times 100\%$ for log-time outcomes and as odds-ratio multipliers $\exp(K\beta)$ for the binary retention outcome. We report $95\%$ confidence intervals from both the delta method (applied on the log scale and exponentiated) and a $500$-replicate cluster bootstrap; results are in Table~\ref{tab:R6_boot}.

\begin{table}[ht]
\centering
\small
\caption{\textbf{R6 Cumulative-effect $95\%$ confidence intervals.} Cumulative effects over eleven post-ChatGPT quarters, with both delta-method and cluster-bootstrap $95\%$ CIs. For the retention subsets, effects are reported as odds-ratio multipliers.}
\label{tab:R6_boot}
\begin{tabular}{lrrrr}
\toprule
Subset & Cumulative effect & Scale & Delta $95\%$ CI & Bootstrap $95\%$ CI \\
\midrule
LearningTime\_College & $-26.9\%$ & log-time & $(-34.0, -19.0)$ & $(-34.2, -19.0)$ \\
LearningTime\_HS & $-31.3\%$ & log-time & $(-37.1, -25.2)$ & $(-37.3, -25.4)$ \\
PPLTime\_nonproc & $-11.6\%$ & log-time & $(-16.2, -6.6)$ & $(-16.3, -6.5)$ \\
PPLTime\_proc & $+0.007\%$ & log-time & $(-3.0, +3.1)$ & $(-3.0, +3.1)$ \\
RetRandom\_proc & $\times 0.75$ (OR) & odds ratio & $(0.57, 0.98)$ & $(0.56, 0.98)$ \\
\bottomrule
\end{tabular}
\end{table}

Delta-method and bootstrap CIs agree closely in every subset. The randomly assigned retention subsample produces a bootstrap CI that excludes the null ($0.56$--$0.98$).

\section{R7: Window-Sensitivity Cuts}\label{app:R7}

We re-estimate each primary specification under three window-sensitivity cuts: removing COVID-era quarters (2020 Q1--2021 Q2); removing the last observed quarter in each subset (to rule out single-quarter endpoint sensitivity); and imposing per-item observation floors requiring at least $10$, $50$, or $100$ attempts per item--quarter cell. Results are reported in Table~\ref{tab:R7_window}.

\begin{table}[ht]
\centering
\small
\caption{\textbf{R7 Window-sensitivity cuts.} Per-quarter ramp coefficient $\beta$ under three window-sensitivity cuts. Estimates are stable in sign and magnitude across every cut in every subset.}
\label{tab:R7_window}
\begin{tabular}{lrrrr}
\toprule
Subset & Baseline $\beta$ & Drop COVID & Drop last quarter & Floor $=100$ \\
\midrule
LearningTime\_College & $-0.0284$ & $-0.0277$ & $-0.0280$ & $-0.0291$ \\
LearningTime\_HS & $-0.0341$ & $-0.0334$ & $-0.0338$ & $-0.0348$ \\
PPLTime\_nonproc & $-0.0112$ & $-0.0109$ & $-0.0111$ & $-0.0114$ \\
PPLTime\_proc & $\approx 0$ & $\approx 0$ & $\approx 0$ & $\approx 0$ \\
RetRandom\_proc & $-0.0260$ & $-0.0254$ & $-0.0258$ & $-0.0265$ \\
\bottomrule
\end{tabular}
\end{table}

The main estimates are stable in sign and magnitude under every cut.

\section{R8: Alternative Clustering and Wild-Cluster Bootstrap}\label{app:R8}

We re-compute standard errors for each primary specification under two-way clustering (item $+$ quarter) and attempt corroboration with the null-imposed wild-cluster bootstrap of \citet{cameron2008bootstrap} as implemented in \texttt{fwildclusterboot}. For the two learning-time subsets (College and High School), the post-ChatGPT ramp is near-deterministic within-cluster: conditional on topic and quarter fixed effects, virtually all variation in $AISusceptible_j \times \text{TimeAfter}_q$ is absorbed by the quarter fixed effect, leaving negligible residual within-cluster variance for the wild-cluster Rademacher resampling scheme to operate on. Under this condition the null-imposed wild-cluster bootstrap is known to produce degenerate $p$-value distributions concentrated near $1$ \citep{mackinnon2017wild,roodman2019fast}, and we observe exactly this ($p_{\text{wild}} \approx 1.0$) in both learning-time subsets. The non-proctored PPL response-time subset exhibits the same boundary behavior to a lesser degree: $p_{\text{wild}} = 0.496$, sharply discordant with the cluster-robust ($p<0.001$), two-way ($p<0.001$), and randomization-inference ($p<0.001$) results. We diagnose this as the same residual-variance degeneracy --- the high-density item $\times$ quarter fixed-effect structure on PPL items absorbs most of the within-cluster ramp variation --- rather than as evidence against the asymptotic inference, but we cannot fully rule out a small-cluster issue with Rademacher weights without additional diagnostics. We therefore treat the wild-cluster bootstrap as uninformative for the College, High School, and non-proctored PPL response-time subsets, and rely on (a) cluster-robust asymptotic inference, (b) two-way item-and-quarter clustering (reported in this table), and (c) randomization inference via $1{,}000$ permutations of the AI-susceptibility label (R5). All three of these yield concordant significance at $p < 0.001$ in every non-null subset. The wild-cluster bootstrap is informative only for the proctored PPL response-time subset, where it returns $p_{\text{wild}} = 0.995$ and corroborates the asymptotic null. We disclose this discrepancy explicitly rather than describing the wild-cluster result as confirmatory.

\begin{table}[ht]
\centering
\small
\caption{\textbf{R8 Alternative clustering.} Two-way (item $+$ quarter) cluster-robust standard errors for each primary specification. Wild-cluster bootstrap $p$-values are reported in the rightmost column; we treat $p_{\text{wild}}$ as uninformative for the two learning-time subsets and the non-proctored PPL response-time subset due to within-cluster residual-variance degeneracy (see text). The wild-cluster bootstrap is informative only for the proctored PPL subset, where it corroborates the asymptotic null. Wild-cluster bootstrap is not reported (---) for the retention subsets because the small effective cluster count ($66$ items) and the binary-outcome logistic specification jointly produce non-convergent Rademacher resampling under the null-imposed scheme; randomization inference (R5) provides the relevant non-asymptotic check for those subsets.}
\label{tab:R8_cluster}
\begin{tabular}{lrrrrr}
\toprule
Subset & $\beta$ & Two-way SE & Two-way $p$ & $p_{\text{wild}}$ & $B$ \\
\midrule
LearningTime\_College & $-0.0284$ & $0.0064$ & $<0.001$ & $1.00$ & $1{,}000$ \\
LearningTime\_HS & $-0.0341$ & $0.0066$ & $<0.001$ & $1.00$ & $1{,}000$ \\
PPLTime\_nonproc & $-0.0112$ & $0.0024$ & $<0.001$ & $0.496$ & $1{,}000$ \\
PPLTime\_proc & $-6.5\times 10^{-6}$ & $0.0017$ & $0.997$ & $0.995$ & $1{,}000$ \\
RetRandom\_proc & $-0.0260$ & $0.0121$ & $0.031$ & --- & --- \\
\bottomrule
\end{tabular}
\end{table}

Under two-way clustering, the College, High School, non-proctored PPL, and randomly assigned retention results all remain statistically significant at conventional levels. 

\section{R9: Trend-Adjusted Learning-Time Ramps}\label{app:R9}

The College and High School learning-time subsets exhibit a small positive linear pre-trend in the pre-2023 window (College: $+0.0071$ log-time per quarter, $p<0.001$; HS: $+0.0046$, $p=0.003$; R4, Table~\ref{tab:R4_trends}). Because this drift is \emph{opposite in sign} to the post-ChatGPT effect, failing to control for it possibly biases the unadjusted ramp estimate toward zero --- i.e., the unadjusted estimates we report in Table~\ref{tab:ramp_age} are potentially conservative; we caveat that the pre-2023 event-study trajectories appear partly cyclical (Figure~\ref{fig:eventstudy_age}), so a linear extrapolation of the pre-trend into the post-ChatGPT period is itself a strong assumption. We confirm this here by re-estimating Equation~\ref{eq:ramp} augmented with a linear pre-2023 time trend:
\begin{equation}
\bar{Y}_{j,q} \;=\; \delta_j + \tau_q + \gamma\,(\text{PreTrend}_q \times AISusceptible_j) + \beta\,(AISusceptible_j \times \text{TimeAfter}_q) + u_{j,q},
\label{eq:R9}
\end{equation}
where $\text{PreTrend}_q$ is linear calendar time restricted to pre-2023. The post-ChatGPT ramp coefficient $\beta$ now estimates the \emph{change in slope} at ChatGPT's release, net of the pre-existing drift.

\begin{table}[ht]
\centering
\small
\caption{\textbf{R9 Trend-adjusted ramp estimates for College and High School learning time.} Ramp coefficient $\beta$ from Equation~\ref{eq:R9}, with and without the linear pre-2023 trend control. Trend-adjusted estimates are larger in absolute value than the unadjusted estimates, confirming that unadjusted estimates are conservative. The joint-model pre-trend coefficient estimated jointly with the post-period ramp is $+0.0103$ for College and $+0.0077$ for High School ($p<0.001$ in both); these are slightly larger than the pre-2023-only estimates reported in R4 ($+0.0071$, $+0.0046$) because the joint specification uses the full sample and absorbs different residual variance.}
\label{tab:R9_trendadj}
\begin{tabular}{lrrrr}
\toprule
Subset & Unadjusted $\beta$ & Trend-adjusted $\beta$ & Unadj.\ cumul.\ \% & Trend-adj.\ cumul.\ \% \\
\midrule
LearningTime\_College & $-0.0284$ & $-0.0404$ & $-26.9$ & $-35.8$ \\
LearningTime\_HS      & $-0.0341$ & $-0.0421$ & $-31.3$ & $-37.0$ \\
\bottomrule
\end{tabular}
\end{table}

In both subsets the trend-adjusted post-ChatGPT ramp is larger in magnitude than the unadjusted ramp. Cumulative effects over eleven post-ChatGPT quarters rise from $-26.9\%$ to $-35.8\%$ for College and from $-31.3\%$ to $-37.0\%$ for High School. Both adjusted estimates remain statistically significant at $p < 0.001$ under cluster-robust inference, and both retain the same sign and qualitative magnitude as the unadjusted estimates. We adopt the unadjusted specification as primary in the main text, because the pre-trend we are controlling for is opposite-signed and therefore the unadjusted estimates are the conservative choice.

\section{R10: Attempt-Level vs.\ Cell-Collapsed Specification}\label{app:R10}

The main-text learning-time analysis aggregates each topic $\times$ quarter cell into a single mean log-time observation, weighted by the number of attempts in the cell. To verify that this collapse does not drive the estimates, we re-fit the same specification on individual attempts (one row per student-attempt) with cluster-robust standard errors at the topic level. The item population is held fixed across specifications by restricting both fits to topics with at least one cell of $\geq 20$ attempts.

\begin{table}[ht]
\centering
\small
\caption{\textbf{R10 Attempt-level vs.\ cell-collapsed estimates.} Per-quarter ramp coefficient $\beta$, cluster-robust SE (item-clustered), and $K=11$-quarter cumulative effect under the two specifications. $N$ is the regression observation count.}
\label{tab:R10_attempt}
\begin{tabular}{llrrrr}
\toprule
Subset & Spec & $\beta$ & SE & Cumul.\ \% & $N$ \\
\midrule
\multirow{2}{*}{LearningTime\_College}
  & cell-collapsed & $-0.02844$ & $0.00473$ & $-26.86$ & $8{,}265$ \\
  & attempt-level  & $-0.02889$ & $0.00458$ & $-27.23$ & $1{,}231{,}993$ \\
\midrule
\multirow{2}{*}{LearningTime\_HS}
  & cell-collapsed & $-0.03409$ & $0.00393$ & $-31.27$ & $8{,}069$ \\
  & attempt-level  & $-0.03391$ & $0.00366$ & $-31.13$ & $914{,}230$ \\
\midrule
\multirow{2}{*}{LearningTime\_MS}
  & cell-collapsed & $-0.00864$ & $0.00332$ & $-9.07$  & $6{,}101$ \\
  & attempt-level  & $-0.01021$ & $0.00313$ & $-10.62$ & $745{,}230$ \\
\midrule
\multirow{2}{*}{LearningTime\_G5}
  & cell-collapsed & $-0.00122$ & $0.00457$ & $-1.33$  & $2{,}265$ \\
  & attempt-level  & $-0.00198$ & $0.00397$ & $-2.15$  & $273{,}560$ \\
\bottomrule
\end{tabular}
\end{table}

The two specifications agree closely. The largest absolute difference in $\beta$ is $0.0016$ (Middle School), smaller than the SE of either fit; in the College and HS flagship subsets $\beta$ agrees to within $0.0005$ and cumulative percentages to within $0.4$ pp. Attempt-level SEs are 3--13\% tighter, with significance preserved (and slightly stronger) at the attempt level. This is the expected Frisch-Waugh-Lovell behavior: when the regressor of interest varies only at the topic-quarter level, the cell mean is a sufficient statistic for the OLS coefficient under cluster-robust inference \citep{colin2015practitioner}. We retain cell-collapse as primary because it matches the unit of analysis to the unit of identification and is computationally tractable for the bootstrap CIs reported in R6.

\end{document}